\documentclass[12pt,preprint]{aastex}

\begin{document}

\shorttitle{Asteroseismology of delta Scuti stars}
\title{Asteroseismology of delta Scuti stars - a parameter study and 
application to seismology of FG Virginis}

\author{Matthew Templeton, Sarbani Basu, and Pierre Demarque}
\affil{Astronomy Department, Yale University, P.O. Box 208101, New Haven, CT
06520-8101} 
\email{mtemplet,basu,demarque@astro.yale.edu}

\begin{abstract}
We assess the potential of asteroseismology for determining the
fundamental properties of individual $\delta$ Scuti stars.  We computed a 
grid of evolution and adiabatic pulsation models using the Yale Rotating
Evolution Code to study the systematic 
changes in low-order ($\ell = 0, 1, 2,$ and $3$) modes as functions of
fundamental stellar properties.  Changes to the stellar mass, chemical 
composition, and convective core overshooting length change the observable
pulsation spectrum significantly.  In general, mass has the strongest 
effect on evolution and on pulsation, followed by the metal abundance.  Changes
to the helium content have very little effect on the frequencies until near
the end of the main sequence.  Changes to each of the four parameters change
the $p$-mode frequencies more, both in absolute and relative terms, than they
do the $g$- and mixed-mode frequencies, suggesting
that these parameters have a greater effect on the outer layers of the star.

We also present evolution 
and pulsation models of the well-studied star FG Virginis,
outlining a possible method of locating favorable models 
in the stellar parameter space based upon a definitive identification of 
only two modes.  Specifically, we plot evolution models on the (period-period 
ratio) and (temperature-period ratio) planes to select candidate models, 
and modify the core overshooting parameter to fit the observed star.  For 
these tests, we adjusted only the mass, helium and metal abundances, and core 
overshooting parameter, but this method can be extended to include the 
effects of first-order rotational splitting and second-order rotational 
distortion of pulsation spectra.

\end{abstract}


\section{Introduction}

Asteroseismology represents the only means to study stellar interiors in
detail.  It uses the observed pulsation spectrum to determine the current 
structural properties of a given star, from which we may deduce the 
evolutionary history of that object.  Several classes of known pulsating 
stars can be studied with this technique, and it raises the possibility 
that we may learn more than ever before about stars other than the Sun.  
The basics of stellar evolution are understood, but many of the finer
details are not yet clear.  Furthermore, many
fundamental properties of stars are difficult or impossible to estimate
observationally (e.g., the helium abundance in low-mass stars).
Pulsations of stars, particularly those in open clusters, may allow us to
finely calibrate fundamental stellar properties including the metal
{\it and} helium abundances and the extent of convective core overshooting.

Currently, $\delta$ Scuti stars are the most promising candidates for 
asteroseismology of stars near the main sequence.  As a class, $\delta$ Scuti 
stars exhibit a wide variety 
of pulsation behavior.  This class includes highly-evolved, high-amplitude,
radial pulsators that are the low-mass cousins of the Cepheid variables,
as well as low-amplitude, radial and non-radial pulsators found on the
pre-main, main-, and post-main-sequences.  
Multi-periodic $\delta$ Scuti stars have been known for decades, and the 
number of independent modes detectable in these stars has increased 
dramatically with improvements in observational technique.  
Some of these objects (e.g., FG Vir: \citet{breger98}; XX Pyx: \citet{hand00};
4 CVn: \citet{breg4cvn}) are known to pulsate in 
dozens of independent modes, and a precise determination of their pulsation
spectra may allow us to determine accurately their interior structures and
evolutionary history.

$\delta$ Scuti stars also happen to be very complex objects.  For one, the 
low-amplitude, main-sequence $\delta$ Scutis tend to be rapid rotators,
with $v \sin{i} > 50$ km/s.  This deforms the spherical symmetry of the star,
and requires a rigorous treatment of rotation and rotational splitting when
modeling these stars \citep{dg92,mich98}.  Some $\delta$ Scuti stars are 
also known to exhibit strong metal lines or have odd line strength ratios, 
suggesting that metal and helium diffusion are also important
both for the static structure and evolution of the star.  $\delta$ Scuti
stars are also massive enough to have a significant convective core during
their main-sequence evolution, which makes core overshooting and mixing
important as well.

Current data are not sufficient to allow an effective inversion of seismic
data to determine the sound speed as a function of depth in these stars, as
can be done for the Sun.  Furthermore, precise seismic inversions require the 
observation of modes with $\ell \geq 3$, something which may not be possible
photometrically with point sources.
As a consequence, forward modeling is still required.  Unfortunately, the
possible parameter space of models required to fit a specific star can be
very large.  We are further hindered by the fact that we often do not know the 
degree of a given mode without a detailed analysis of narrow-band photometry 
or spectral line equivalent widths \citep{viskum97}.  However, observations are
nearly at the point where precision asteroseismology can be done on these
sources.  Furthermore, new space-based photometry missions may detect 
additional, low-amplitude modes beyond what is currently detectable with 
ground-based observations.  This may allow us to perform analyses of large 
and small frequency separations \citep{gn90} and 
$n$-th differences \citep{gough90,basu97}
which may allow us to estimate the stellar age and determine the sizes of 
the convection zones.

An understanding of how pulsation behavior changes as a function of age, mass, 
composition, temperature, and the effects of convective overshoot is required
to constrain more effectively observed stars with theoretical models.
In this paper, we explore how these first-order parameters affect the 
pulsation spectra of these models.  This work will set limits on the 
usefulness of pulsation spectra for fitting stellar models.  It will also 
provide a basis for expanding the discussion of model fitting to involve 
second order effects like rotation and diffusion.  We also apply this 
information to modeling of the $\delta$ Scuti star FG Virginis, and present
a model obtained using this analysis.

The first half of the paper will show how stellar evolution and pulsation
behaviors change as functions of the stellar mass, chemical composition, and
convective core overshooting parameters.  In particular, we are interested in
quantifying the frequency shifts caused by changes in these parameters, in 
hopes that we may use this information to better determine stellar properties
in the absence of information (e.g., binary masses, spectroscopic metal 
abundances, temperatures, and surface gravities).  We also discuss useful
seismological diagnostics including period ratios and frequency separations,
with an emphasis on possible applications of space-based photometric data.
The second half of the paper will apply the discussion of the first half to
the fitting of models to a specific star, the well-studied multiply-periodic
$\delta$ Scuti star FG Virginis.  We present our method of model fitting
using period ratios of identified modes along with the photometrically derived
effective temperature, and present a best-fit model.  We briefly discuss the
effect of rapid rotation on the overall results of the paper.

\section{Evolution models}
The stellar models we use in this work were generated with the Yale Rotating
Evolution Code \citep{yrec}, hereafter YREC.  The evolution models were 
calculated using the 
OPAL \citep{opale} equation of state.  The nuclear reaction rates were 
supplied by J. Bahcall and M. Pinsonneault.  OPAL \citep{opalo} and \citet{a95}
opacities were used at $\log{T} > 4.1$ and $\log{T} \le 4.1$
respectively.  Rotation and diffusion effects were not included in the
evolution calculations, but convective core overshooting was included in
specific models noted below.  For convection, we used the standard mixing
length approximation, with a mixing length parameter $\alpha = 1.92$.  This
value yielded the closest match between the YREC models and models
computed previously with the Iben evolution code \citep{tem01,tbg00}.
All models were made with chemical compositions using the \citet{gn93}
solar metal abundance ratios, and an initial
He$^{3}$ mass fraction of $2.9 \cdot 10^{-5}$.

\begin{figure}
\plotone{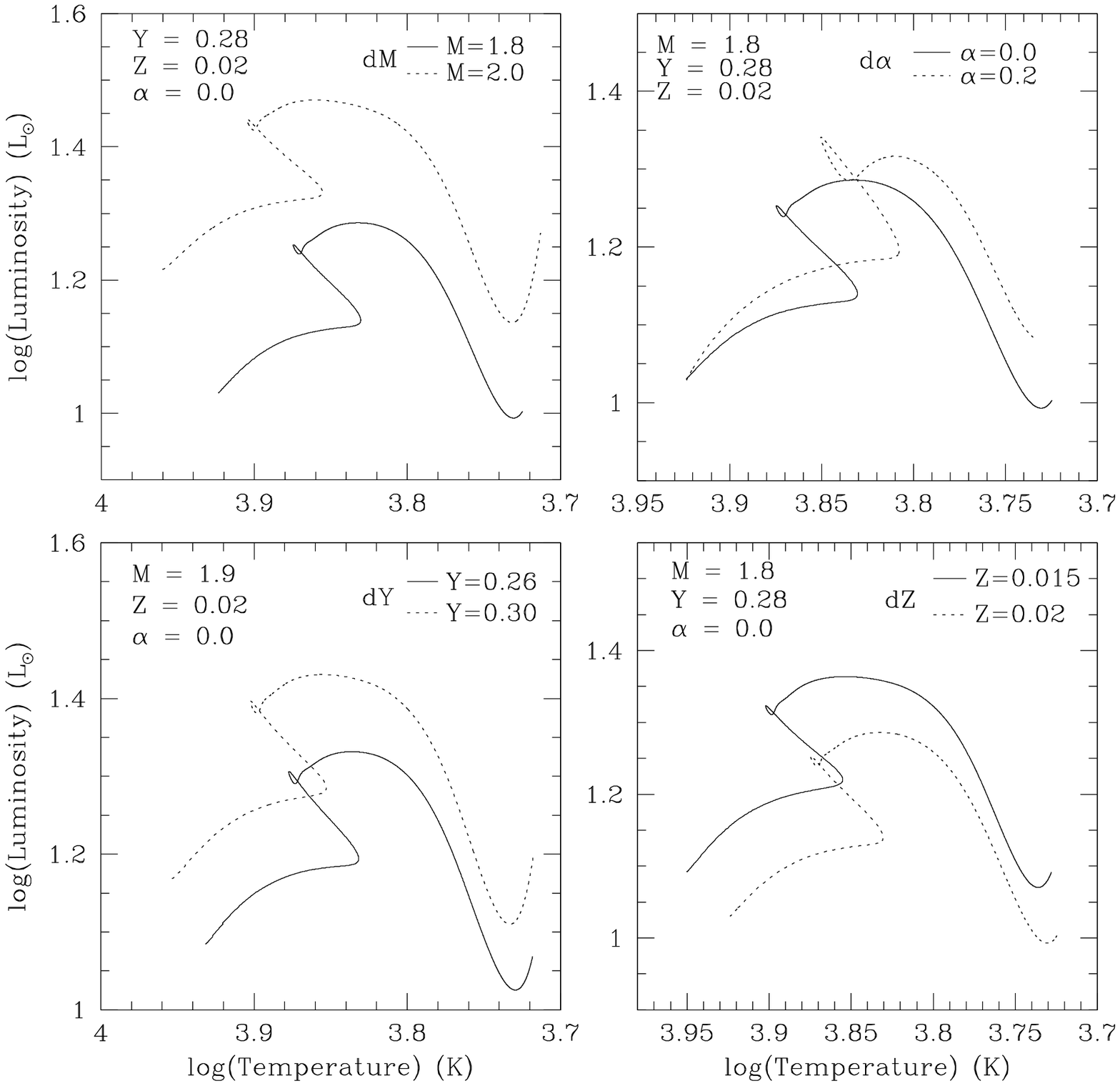}
\caption{Evolution tracks of the four pairs of models, showing how modifying
each of the fundamental parameters effects the model.  Changes to the mass
and chemical composition shift the entire evolution track diagonally in the
temperature-luminosity plane.  Changes to the overshooting parameter change
the evolution track with time, though the ZAMS location is unchanged.}
\label{fig:one}
\end{figure}

Four pairs of models were evolved.  In each pair, one parameter (mass, helium
abundance, metal abundance, or convective core overshooting parameter) was
changed while the other three were held fixed.  In this way, we can test the
effects of varying one single parameter on both the evolution and pulsation
behavior.  The evolution tracks are shown in Figure \ref{fig:one}, and their 
properties
listed in Table \ref{tab:one}.  All models were evolved through the post-main 
sequence to
near the base of the red giant branch (down to at least 6000 K).  Every 20
evolution time steps, we generated pulsation models (with envelopes and
atmospheres) to study how the pulsation frequencies vary as a function of age.
These pulsation models will be discussed in the next section.

\begin{table}[htbp]
\caption{Evolution model initial conditions for the grid. $Y_{0}$ and $Z_{0}$
are the initial helium and metal mass fractions, and $\alpha_{C}$ is the
convective core overshooting parameter in pressure scale heights.  Models
without overshoot are labeled ``none''.}
\label{tab:one}
\begin{center}
\begin{tabular}{cccc|cccc}\hline
$Z_{0}$ & $Y_{0}$ & Mass & $\alpha_{C}$ & $Z_{0}$ & $Y_{0}$ & Mass & 
$\alpha_{C}$ \\
\hline \hline
0.02 & 0.28 & 1.8 & none  & 0.02 & 0.28 & 1.8 & none\\
0.02 & 0.28 & 1.8 & 0.2   & 0.02 & 0.28 & 2.0 & none\\
\\
0.02 & 0.26 & 1.9 & none  & 0.015 & 0.28 & 1.8 & none\\
0.02 & 0.30 & 1.9 & none  & 0.02 & 0.28 & 1.8 & none\\
\hline
\end{tabular}
\end{center}
\end{table}

\subsection{The behavior of evolution models with changes to initial parameters}
The behavior of the evolution models with changes to fundamental parameters
was as expected, based on previous studies.
Pairs of models with different masses and initial chemical compositions have
significantly different ZAMS locations.  Increasing the mass and helium 
abundance results in higher luminosities and effective temperatures, while
decreasing the metal abundance has the same effect.  Increasing the mass by
$0.2 M_{\odot}$ increased the luminosity of the ZAMS point by 0.2 dex, and 
the temperature of the ZAMS by 0.04 dex; the luminosity difference is
maintained throughout the pre-red giant branch life of the star, though the
main sequence of the higher mass star has a wider range of temperature.  
Increasing the helium abundance has the same effect: increasing $Y$ from 0.26 to
0.30 increases the ZAMS luminosity by 0.1 dex, and the temperature by 0.02 dex,
again maintained through the life of the star.  {\it Decreasing} the metal
abundance $Z$ has the same effect.  Reducing the metal mass fraction from 
0.02 to
0.015 increased the ZAMS luminosity by slightly less than 0.1 dex, and the 
ZAMS temperature by 0.02 dex.  The reduction in CNO cycle luminosity caused
by the drop in metal abundance is offset by the reduction in photospheric
opacity, yielding a hotter and more luminous star.  

A change in convective core overshooting produces a fundamentally different
effect than changes in the other three parameters; there is no difference in 
the ZAMS
locations of the overshooting and non-overshooting models, since changes in
core size and composition take time to appear.  The overshooting model is
more luminous at a given temperature, and the main sequence of the
overshooting model is longer (by 300 Myr) due to the addition of unburned
material to the core, and the turn-off occurs at a lower temperature.  

Of the four parameters, metal abundance is the only one directly detectable
via spectroscopy or precision photometry.
It is difficult to precisely determine the mass of a given star
unless it is a member of a binary or a cluster.  Str\"{o}mgren
colors can provide good estimates of temperature and surface gravity, but 
the accuracy in temperature is usually not better than a few hundred kelvins
and the surface gravity about 0.1 dex in $\log{g}$ \citep{nsw93}.
{\em HIPPARCOS} \citep{hip97} and the {\em HST} Fine Guidance Sensors 
\citep{har00} have been used to determine parallaxes and hence luminosities, 
but again, these are not accurate to better than ten percent.  The 
overshooting parameter is impossible to calibrate photometrically without much 
more accurate determinations of the mass, temperature, and luminosity for 
individual stars.  The helium content of individual stars (cooler than 
about 20,000 K) represents an even more difficult problem, as the effects of 
increasing and decreasing the helium content mimic the temperature and 
luminosity changes from increasing and decreasing the mass, particularly while
the star is still on the main sequence.  Therefore, while photometric
methods can be used to estimate stellar properties, seismology provides the
only reasonable hope of precisely determining them.

\section{Pulsation behavior}
In this section we study how the pulsation mode frequencies are affected by
changes to the input model's fundamental parameters.  We compare sets of model
pulsation frequencies where we change the mass, chemical composition, and
core overshooting parameter individually, along with a change in age as the
star evolves.  

To determine the pulsation frequencies, we used the pulsation models generated
by the YREC program as inputs for a simple adiabatic pulsation code.  Each
model typically had about 500 zones in the core, 1000 in the envelope, and
2000 in the atmosphere.  We computed the pulsation frequencies for 
$\ell = 0,1,2,3$ modes between 40 and 4000 $\mu$Hz.  The upper limit of
4000 $\mu$Hz is much
higher than frequencies currently observed in $\delta$ Scuti stars, and for
much of the evolution sequence, is much higher than the acoustic cutoff 
frequency.  However, we wanted to study the evolution of all of the frequencies
as a function of time, particularly since space-based photometry missions may
be able to detect high-frequency $p$-modes.  The acoustic cutoff frequency also
evolves as a function of time, and we wanted to obtain all the frequencies
theoretically permitted in each model, without having to compute the acoustic
cutoff for each case in advance.  We did not compute the nonadiabatic 
frequencies because the nonadiabatic corrections to the frequencies are small,
and because we are not as concerned with whether frequencies are excited 
as we are with the frequencies themselves.

\subsection{The model pulsation spectra}
\begin{figure}
\plotone{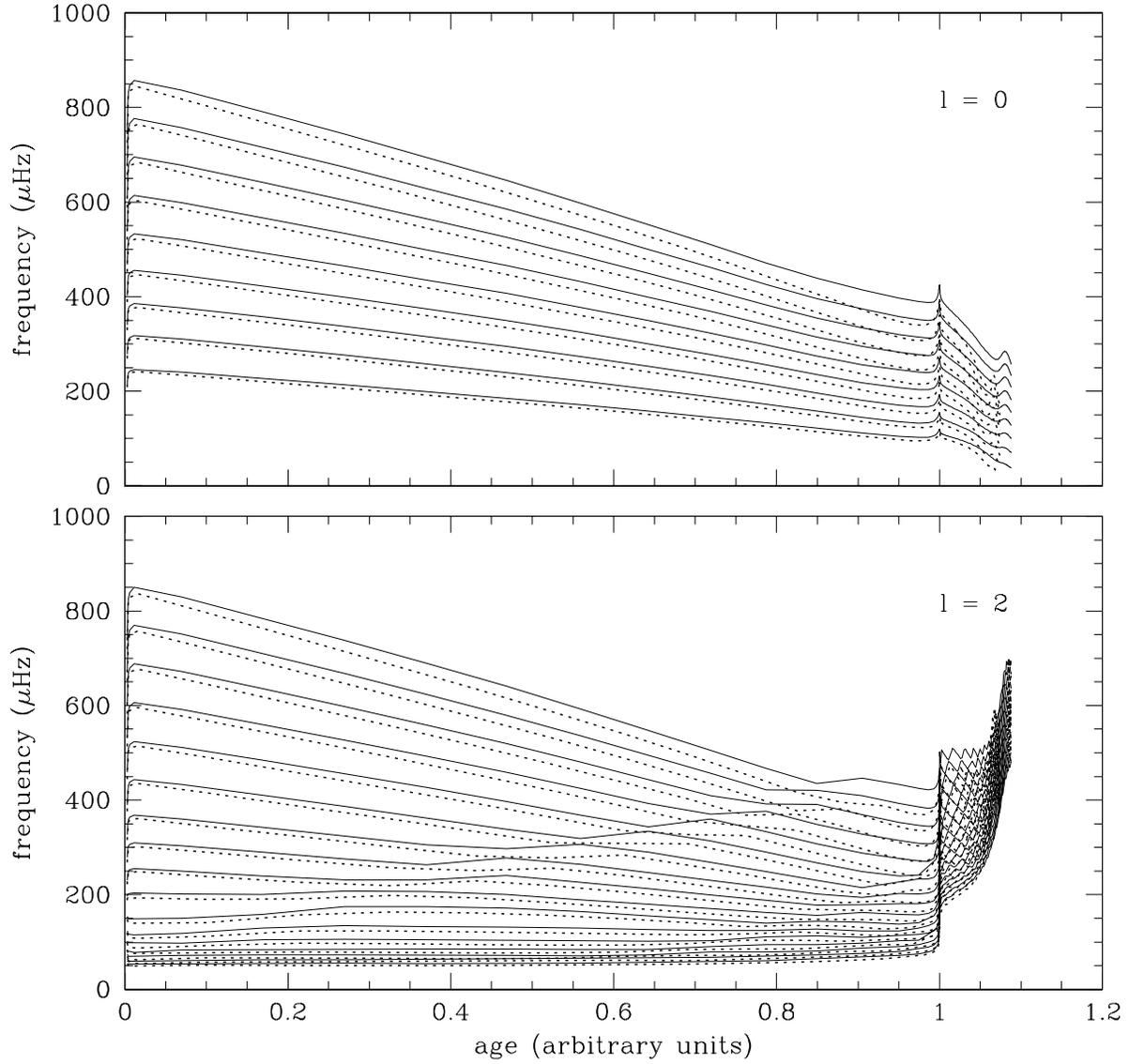}
\caption{Evolution of the pulsation spectra of two models with different
masses, having identical
initial chemical compositions $Y$ and $Z$, and no convective core overshoot.
Solid lines -- $M = 1.8 M_{\odot}$; dotted lines -- $M = 2.0_{\odot}$.  Model
ages have been normalized so that the frequency spike near the end of core
hydrogen burning has a normalized age of 1.  The actual ages of the models
at this point are 1.2245 and 0.90862 Gyr, respectively.}
\label{fig:two}
\end{figure}

\begin{figure}
\plotone{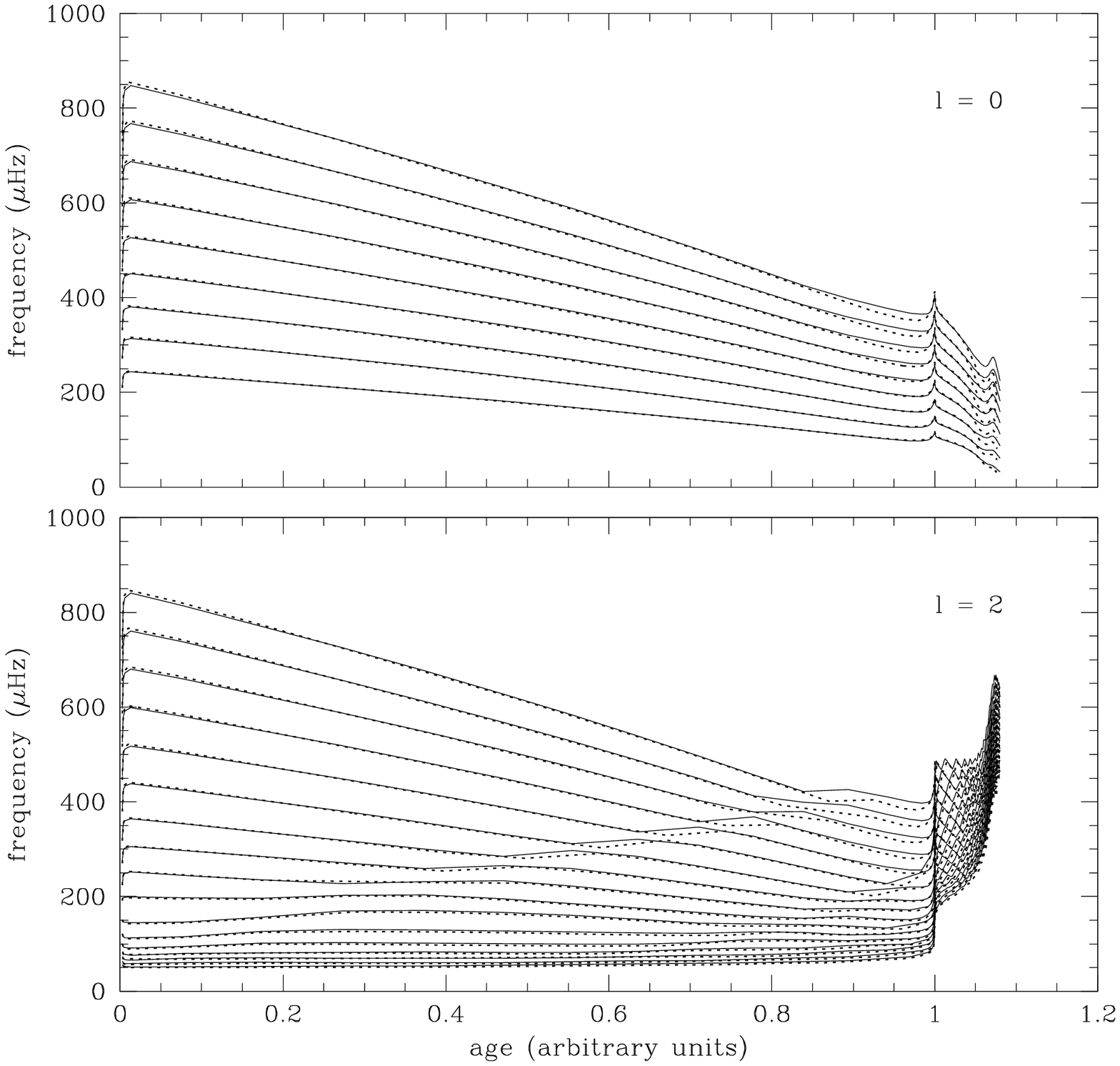}
\caption{Evolution of the pulsation spectra of two models with different 
initial helium abundances $Y$, having identical masses and initial metal 
abundances $Z$, and no convective core overshoot.  Solid lines -- $Y = 0.26$;
dotted lines -- $Y = 0.30$.  Model ages have been normalized so that the 
frequency spike near the end of core hydrogen burning has a normalized age 
of 1.  The actual ages of the models at this point are 1.1764 and 0.94106 
Gyr, respectively.}
\label{fig:three}
\end{figure}

\begin{figure}
\plotone{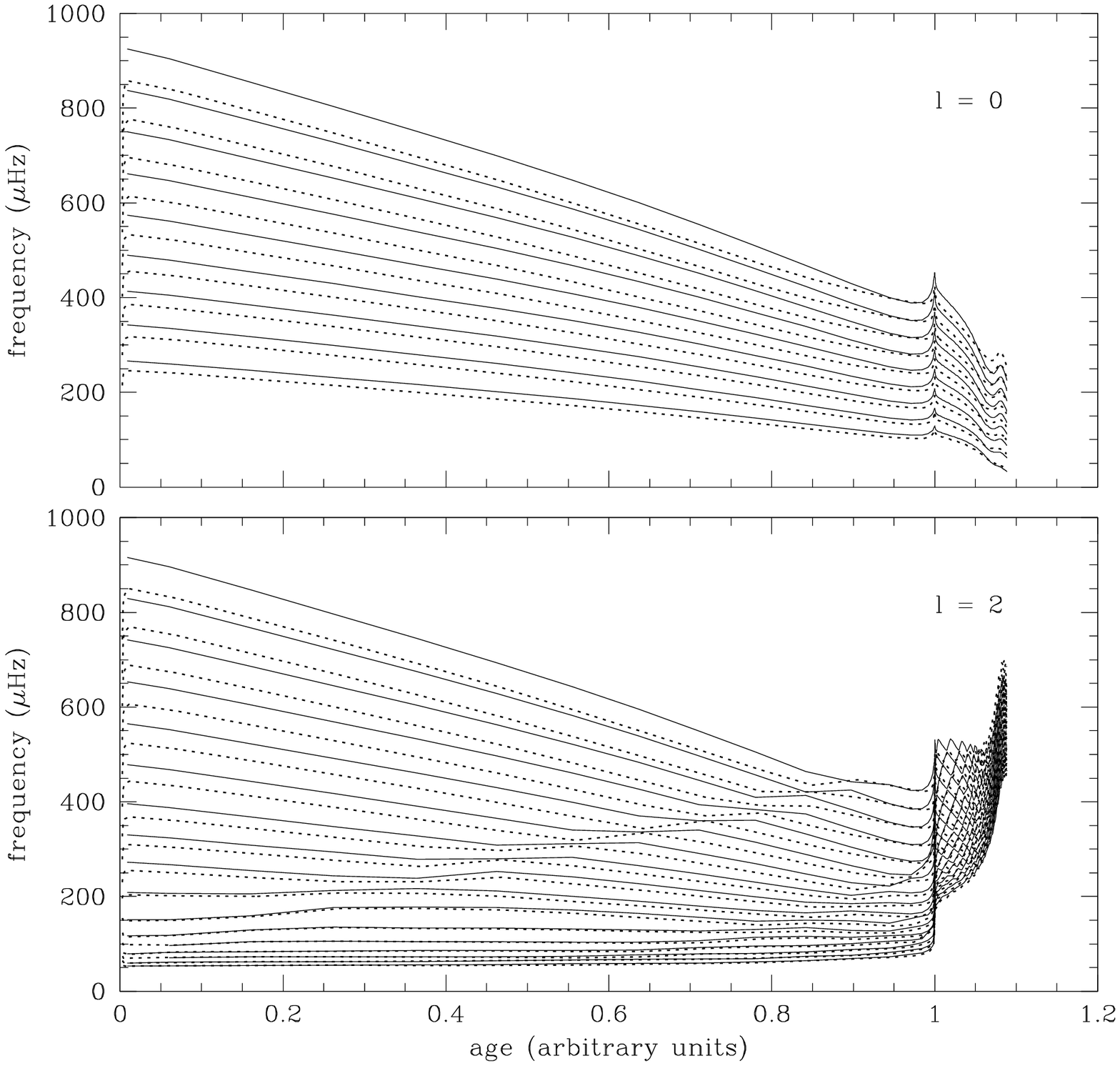}
\caption{Evolution of the pulsation spectra of two models with different
initial metal abundances $Z$, having identical masses and initial helium
abundances $Y$, and no convective core overshoot.  Solid lines -- $Z = 0.015$;
dotted lines -- $Z = 0.02$.  Model ages have been normalized so that the 
frequency spike near the end of core hydrogen burning has a normalized age 
of 1.  The actual ages of the models at this point are 1.0878 and 1.2245 
Gyr, respectively.}
\label{fig:four}
\end{figure}

\begin{figure}
\plotone{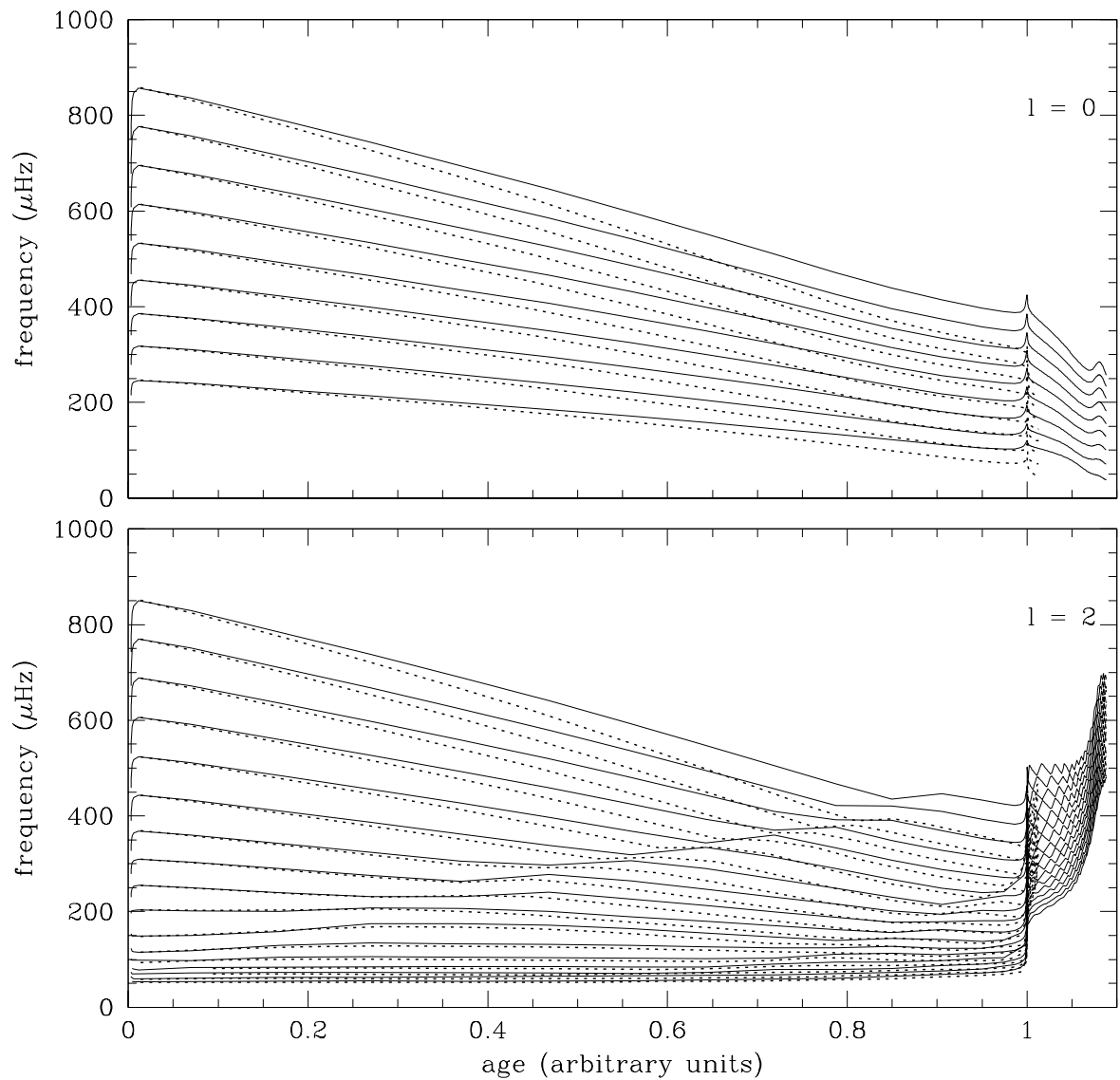}
\caption{Evolution of the pulsation spectra of two models with different
convective core overshooting parameters $\alpha$, having identical masses
and initial chemical compositions.  Solid lines -- $\alpha = 0.0$ (no
overshooting); dotted lines -- $\alpha = 0.2$.  Model ages have been normalized
so that the frequency spike near the end of core hydrogen burning has a 
normalized age of 1.  The actual ages of the models at this point are 1.2245
and 1.5263 Gyr, respectively.}
\label{fig:five}
\end{figure}

Figures \ref{fig:two}, \ref{fig:three}, \ref{fig:four}, and \ref{fig:five} 
show the evolution of $\ell = 0 (0 \geq n \geq 8)$
and $\ell = 2 (-8 \geq n \geq 8)$ modes for pairs of models with different
masses, initial helium abundances, initial metal abundances, and convective
core overshooting parameters respectively.  (Here, $n$ is defined as the number
of $p$-type nodes minus the number of $g$-type nodes for each mode.)  In 
each plot, we normalize age such
that the point on the early post-main-sequence where the frequencies reach
a maximum is equal to one.  With the ages normalized, we can discuss how their 
{\it relative} ages compare.  We note that for all models tested, the
$p$-mode (high) frequencies decay with increasing age, because they are
dominated by the sound crossing time in the envelope, which is in turn
dominated by the (slowly increasing) stellar radius.  The $g$-mode frequencies
remain nearly constant over time, with drastic changes occurring only when
the core structure changes dramatically, as near the end of the main-sequence.
The mode frequencies for these models all have a brief spike at the point
of the main-sequence turn-off caused by a brief period where the temperature
increases during core contraction.  After this point, the radial mode
frequencies decline again, and the $g$- and mixed-mode frequencies increase.

The non-radial modes behave differently than the radial ones, primarily 
because they can show mixed $p$- and $g$-mode character, and because they are
susceptible to avoided crossings \citep{osaki75,cd2000}.  The non-radial 
$p$-mode behavior is very similar to that of the radial modes until they are 
``bumped'' to higher frequencies by $g$- or mixed modes, and take on a 
mixed-mode character of their own.  This bumping does not occur until some 
time on the main sequence has elapsed, and the $g$-mode frequencies approach 
those of the n=1 $p$-mode (in this case, around 0.4 Gyr).  The mode-bumping 
then progresses to higher and higher frequencies
as the dense core evolves and allows more $g$-modes to propagate.  All of the 
mode orders behave the same way, and the bumping occurs at nearly the same
time for all $\ell$-values.

\subsubsection{Changes in mass}
Figure \ref{fig:two} shows the pulsation frequency evolution of a pair of 
models with 
$Z = 0.02, Y = 0.28$, and $\alpha = 0.0$ (no convective core overshooting)
having masses of 1.8 and 2.0 $M_{\odot}$.  Changing the mass has the most
noticeable effect on the pulsation frequencies among all of the parameters
tested.  The two primary differences in the pulsation behavior are that the
more massive star has lower frequencies overall for both $p$- and $g$-modes, 
and that mode-bumping (avoided crossings) occur closer to the main sequence
turn-off point.  The latter point is consistent with a larger convective
core taking longer to evolve.  The radial modes of the higher-mass star
have a steeper decline in frequency, particularly the higher-order modes
($n > 2$) which yield higher radial frequency ratios.

The lower-order $p$- and $g$-type modes maintain a nearly constant frequency
shift with age.  The frequency difference between the two models is small
but measurable in the $\ell = 0$ $p$-modes, with frequency differences of
order 8-15 microhertz.  The frequency differences become more severe at higher
$n$, due to the growing influence of the outer envelope at higher radial 
orders.  The $\ell = 2$ modes show similar behavior.  The higher-order
$g$-modes ($-8 \geq n \geq -4$) show differences of about 5 microhertz, progressing
to near 10 microhertz at $n \sim 0$.  The behavior of the $p$-modes is similar
to that of the $\ell = 0$ modes.  The fact that the large frequency 
separations ($\delta \nu = \nu_{n} - \nu_{n-1}$) of the two models are not
constant with increasing $n$ means that the mean separations between the
models will also differ, though only slightly.  This may help in the case
where specific mode identifications are not known, only an average frequency
separation.

\subsubsection{Changes in initial helium abundance}
Figure \ref{fig:three} shows the pulsation frequency evolution of a pair of 
models with 
$M = 1.9 M_{\odot}$, $Z = 0.02$, and $\alpha = 0.0$ having initial helium 
abundances of $Y = 0.26$ and $Y = 0.30$.  There is very little difference
in the pulsation frequencies of the two models, until nearly eighty percent
of the main-sequence lifetime of each has elapsed.  For the low-order 
$\ell = 0$ modes, the frequency difference between the models is very small,
less than one microhertz, over most of the main sequence.  At higher orders, the
difference between the two models becomes more pronounced, again because of
differences in the outer envelope.  Differences are also small in the 
$\ell = 2$ modes, though in this case, the $g$-modes show a larger difference,
on the order of two or three microhertz.  

The small difference between the two models is surprising given
that the evolution tracks for these stars are so different (the model with
the higher initial helium abundance is about 500 K hotter and twenty percent
more luminous over its lifetime).  It suggests that the increase in radius of
the helium-rich star is offset by a higher mean sound speed throughout the
star.  The similarity of the $g$-modes and the ages at which the non-radial
mode bumpings occur suggests that the pace of core evolution is similar in
both cases.  The overall similarity of frequencies suggests that the pulsation
frequencies are not a good diagnostic of the helium abundance, and that
the luminosity and effective temperature would be better ones.

\subsubsection{Changes in initial metal abundance}
Figure \ref{fig:four} shows the pulsation frequency evolution of a pair of 
models with
$M = 1.8 M_{\odot}$, $Y = 0.28$, and $\alpha = 0.0$ having initial metal
abundances of $Z = 0.015$ and $Z = 0.02$.  This case is intermediate between
that of changing the mass and changing the helium abundance.  The
$p$-modes appear to be the most affected.  At lower frequencies, the
evolution of the $\ell = 0$ $p$-modes closely parallel one another, with the
higher-metal abundance models maintaining a nearly constant (lower) frequency
offset.  Additionally, there is a significant offset in frequencies, nearly
20 microhertz on the main sequence.  The frequencies of the higher-order 
$\ell = 0$ modes drop more rapidly as time progresses, which results in higher
frequency ratios between the radial modes.  The non-radial $p$-modes show 
similar behavior.  On the other hand, the lower-frequency $g$- and mixed-modes 
($n \leq -5$) appear to be essentially unchanged.  Differences between the
two models' $g$-mode frequencies are very small, less than one microhertz over 
most of the main sequence lifetime.  This suggests that the $g$-mode frequencies
are a poor diagnostic of the metal abundance, at least in the early portion
of the main sequence.  

\subsubsection{Changes in convective core overshooting parameter}
Figure \ref{fig:five} shows the final case of the pulsation frequency evolution
for a pair
of models with $M = 1.8 M_{\odot}$, $Y = 0.28$, and $Z = 0.02$ having
convective core overshooting lengths of $\alpha = 0.0$ (no overshooting) and
$\alpha = 0.2$ (substantial overshooting) pressure scale heights.  As would
be expected, the frequencies are identical at the ZAMS (age = 0) but start
to deviate strongly thereafter.  The high-frequency $p$-modes show the strongest
deviations with age due to their strong dependence upon the stellar radius, 
but eventually all of the frequencies in the overshooting model are 
systematically lower.  The high-order $p$-modes differ by as much as 50 
microhertz
between the two models, dropping to 10-20 microhertz in the low-order $p$-modes,
and to 5-10 microhertz for the $g$-modes.  The $p$-mode differences most likely
occur due to the much lower temperatures reached by the overshooting model
at the end of the main sequence, while the $g$-modes will be more influenced
by the larger core.

The behavior of the overall spectrum is also interesting.  Mode-bumpings 
occur (relatively) earlier in the overshooting model due to the faster drop 
in the $p$-mode frequencies.  This is somewhat surprising, since one could 
assume that an overshooting core will evolve less quickly and would not force 
the $g$-mode frequencies upwards as soon.  It is noteworthy also that the 
post-main-sequence evolution between the end of the main-sequence and the 
base of the red giant branch is
more rapid in the overshooting model, seen in the very sharp drop in the
$\ell = 0$ frequencies immediately after the frequency spike.  This is
presumably due to the partial depletion of hydrogen from the shell-burning
region of the star, resulting in a more rapid drop to the base of the red
giant branch.  This raises an interesting possibility, namely that the period
change ($\dot{\Pi}$) of a star with convective core overshooting should be 
measurably faster than that of a star without 
overshooting.  This might provide a way of calibrating the convective core
overshooting parameter using the period changes observed in some 
post-main sequence $\delta$
Scuti stars.

\subsubsection{Observational consequences}
\begin{figure}
\plotone{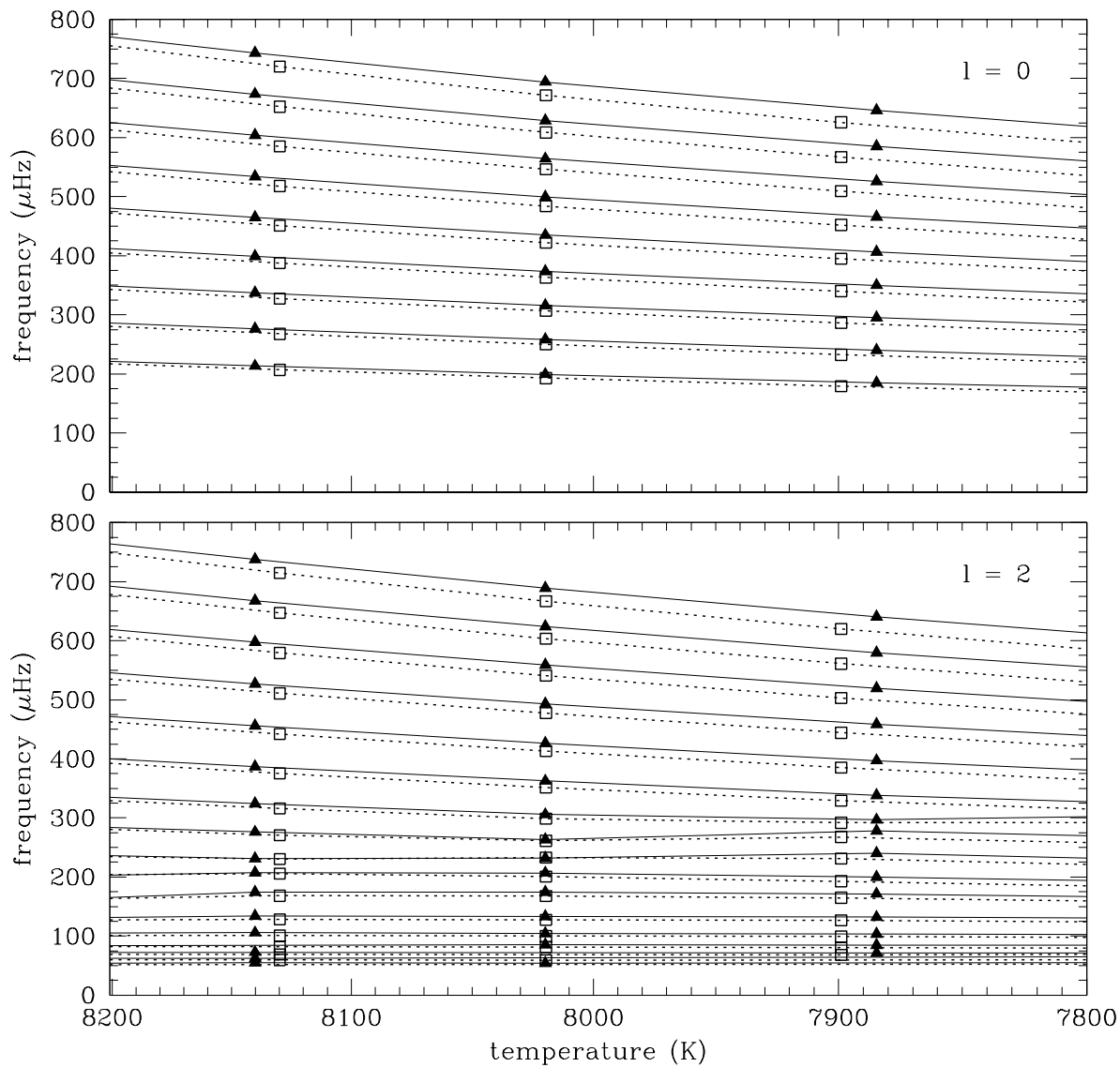}
\caption{Evolution of the two models with different convective core 
overshooting parameters, plotted as a function of temperature rather than
age.  Solid lines and triangles -- $\alpha = 0.0$ (no overshoot); 
dotted lines and open squares -- $\alpha = 0.2$.  In the temperature-period
plane, differences between models appear smaller.  In the absence of rotation,
these differences would be measurable, but rotational splitting could be
large enough to mask the differences.}
\label{fig:six}
\end{figure}

The analysis above is useful for seeing how the star evolves over time, and
in general, the differences in frequency evolution are clear.  However,
the problem with this is that the age itself is not an
observable quantity.  The primary observational parameter for these stars is
the effective temperature, with the luminosity and surface gravity being
secondary.  When the two different models are compared at the same temperature
rather than the same evolutionary stage, the differences are still visible
but harder to see.  Figure \ref{fig:six} shows the evolution of the models in
Figure \ref{fig:five}, but focuses on a narrow range in temperature.  At a
given temperature, the differences in overshooting parameter yield a modest
but visible differences in the frequencies of between five and ten
microhertz, particularly in the higher frequency $p$-modes.  These differences
are well within the observational error obtained with multi-season photometric
data.  The main difficulty then is with mode identification, particularly since
the number of modes theoretically predicted in a given star is much higher
than the number of modes actually observed, and the mechanism by which modes
are excited is not well understood \citep{dk90}.  Photometric mode 
identifications based upon temporal phase differences between filters
are possible, and have been used in several stars (e.g., $\theta$ Tuc:
\citet{ps00}; FG Vir: \citet{bppg99}), so this problem could
potentially be overcome.

There is an important complication to the above analysis, namely the problem
of rotation.  All of the models above were tested without the inclusion of 
rotation, either in the evolution calculations or in the pulsation 
calculations.  $\delta$ Scuti stars, particularly those still on the main 
sequence, are known to have relatively high rotation velocities -- 
$v \sin{i} \geq 50$ km/s are not uncommon.  In such cases, one must include the 
effects of rotational splitting of the $\ell > 0$ modes, as well as the 
second and third order effects of rotational perturbation of stellar 
structure \citep{sgd98}.  First, first-order rotational splitting can easily 
reach the level of 5 microhertz for higher rotational velocities, particularly 
if $\ell \geq 2$.  Second, second- and third-order rotation will shift all 
frequencies of a given multiplet by a significant amount, further complicating
the frequency changes due to changes in the fundamental parameters mentioned 
above.  Rotation will also couple modes of $\ell = 0,2$ and $1,3$, which causes
an additional frequency perturbation, and will result in ambiguous mode
identifications.
%Rotation also raises an additional problem, namely mode-coupling.  
%\citet{sgd98} also showed that modes with $\ell = 0$ and 2 (and $\ell = 1$ 
%and 3) with similar frequencies are coupled in the presence of rotation,
%which can shift the individual mode frequencies and make mode identification
%impossible.

In this work, we avoid the problem of rotation entirely, except for the
inclusion of first-order rotation in Section 4.  We feel this is justified
because ultimately what we are doing is a study of the effects of
{\it individual} parameters.  Rotation is an additional parameter just like
the mass, chemical composition, and overshooting length.  But unlike the other
parameters, rotation can be much more complicated when we also include it in
the evolution calculations, and when we also study differential rotation.  We
therefore leave the study of rotation for a later paper, with the caveat that
rotation {\it is} an important additional parameter, and must be included
in detailed asteroseismology of individual stars.

\subsection{Other diagnostics: frequency pairs and ratios}
Frequency ratios of radial modes have been used for quite some time as a
diagnostic for the high-amplitude $\delta$ Scuti stars (HADS), as well as the
Cepheids \citep{pet73,ckh79}, and were even used to test the accuracy of the 
OPAL opacities by solving the frequency ratio discrepancy between predicted 
and observed ratios in the Cepheids \citep{pet93}.  They are useful 
diagnostics for these stars 
because while HADS and Cepheids usually pulsate in only a few modes, the
frequency ratio is sensitive to the fundamental parameters of the star.
For stars with few identifiable modes, frequency pairs are also a convenient
method of constraining the parameter space of possible model fits.  Finally,
pairs of radial $\ell = 0$ modes are not subject to rotational splitting, and
are only subject to relatively small shifts due to higher-order rotation
effects, and are thus easier to fit than the nonradial modes.  As in the
previous section, we again note that rotational mode-coupling may complicate 
matters with the radial modes.  In the high-amplitude stars where double-mode
pulsation is commonly seen, the rotational velocities are generally much 
lower \citep{sf97}, so mode-coupling may not be as great a difficulty for
the more evolved high-amplitude stars.

\begin{figure}
\plotone{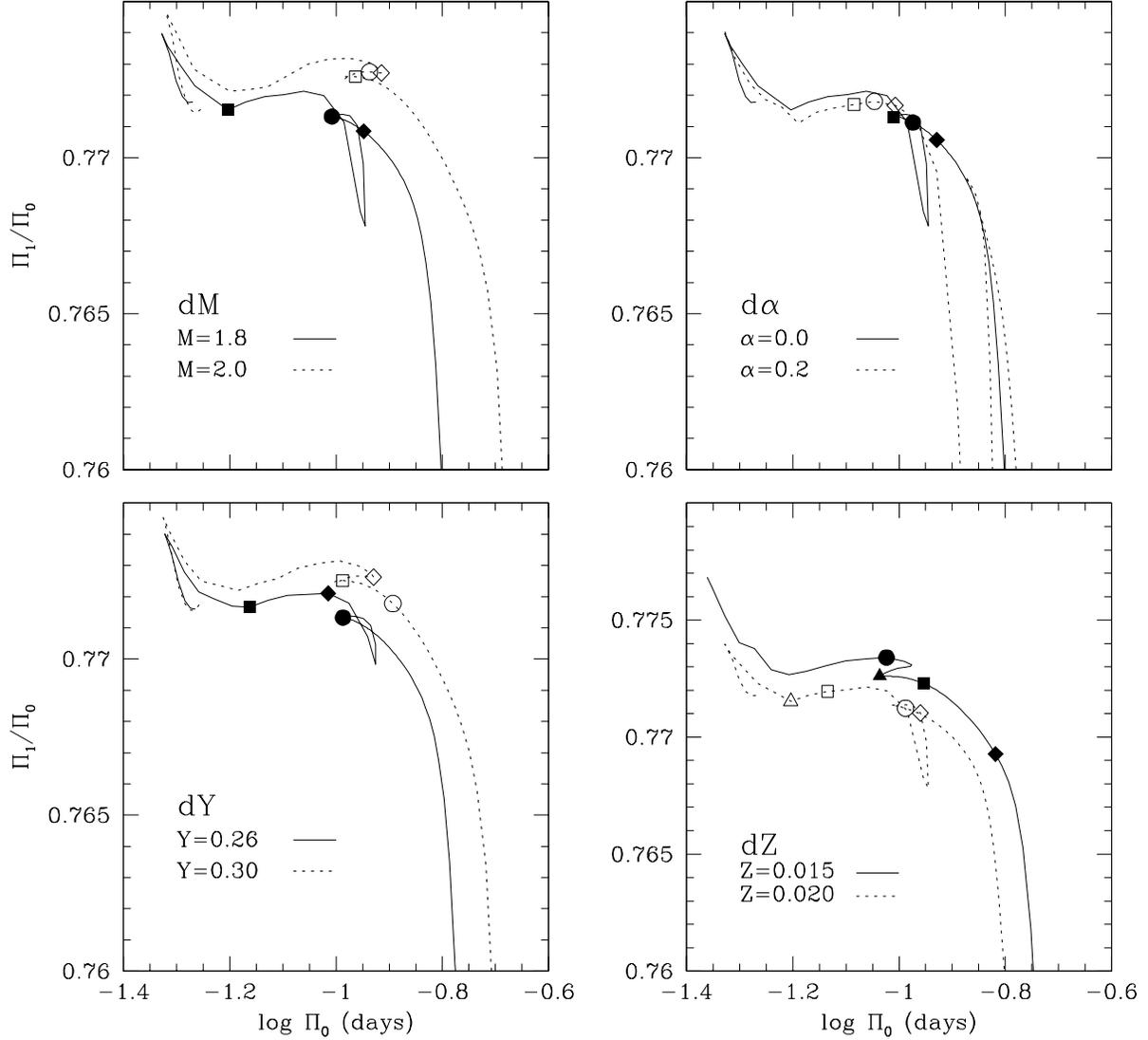}
\caption{Petersen diagrams showing the evolution of the fundamental period
$\log{\Pi_{0}}$ versus the ratio of the first two radial periods.  Stellar
parameters are the same as in Figure 1.  Points along the evolution track
represent locations of equal temperature.  
$dM$: squares -- 7885 K, circles -- 7446 K, diamonds -- 7202 K.
$d\alpha$: squares -- 7470 K, circles -- 7303 K, diamonds -- 7130 K.  
$dY$: squares -- 7870 K, circles -- 7489 K, diamonds -- 7154 K.
$dZ$: triangles -- 7882 K, squares -- 7573 K, circles -- 7359 K, 
diamonds -- 6910 K.}
\label{fig:seven}
\end{figure}

Figure \ref{fig:seven} shows the Petersen diagrams (the evolution of radial 
period ratio versus the log of the fundamental period) for the four pairs of 
evolution models shown in Figure \ref{fig:one}.  In each pair, there are 
measurable differences between the two models; these differences are easily 
detectable despite the period ratios differing by a small value.  Like the 
raw frequency spectra, the period ratios of the different models 
diverge the most when the end of the main sequence is reached 
(near $\log{\Pi_{0}} = -1.0$ in the figures), and the models with different 
masses again appear to differ the most.  The models with different helium and 
metal abundances also show differences throughout the evolution track.
The models with different core overshooting parameters show a smaller 
difference on the main sequence, but diverge drastically at the main sequence 
turn-off.  The two models that have relatively large excursions to lower
period ratios are the $M = 1.8 M_{\odot}$, $Y = 0.28$, $Z = 0.02$ models with
and without overshoot.  The primary reason for this is that at the 
main-sequence turn-off, the temperature of these models drops to the point
where a significant near-surface convection zone is established.  This acts
to lower the period ratio.  The overshooting model moves to cooler 
temperatures, and thus drops to lower period ratios.  A similar effect is
seen at smaller masses with cooler turn-off points.

The period-period ratio diagram is useful when the only information available
are the periods.  If the temperature can also be determined, it provides a 
useful additional constraint.  The points in Figure \ref{fig:seven} denote 
locations
of equal temperature in both models.  It is important to note that models
at the same temperature can have very different fundamental periods and period
ratios.  This is particularly evident in the d$\alpha$ and dY plots, and for 
a few points in the $dM$ and $dZ$ plots.  Conversely, models with similar 
periods may have very different effective temperatures, thus placing a 
further constraint on the model parameter space.  However, in a few cases,
the fundamental period and effective temperature of the two models are
similar (within 10 microhertz), and it is only the period ratio that distinguishes
one from the other.  Similarly, the effective temperature might be useful to
distinguish between models whose evolution tracks of period and period ratio 
intersect.

\subsection{Other possible diagnostics}
Upcoming space-based photometry missions may detect many of the unobserved 
pulsation modes in $\delta$ Scuti star spectra.  If a majority of the 
theoretically predicted modes are observed, then other seismological 
diagnostics may become useful.  The simplest of these, the large separation
mentioned above, has already been measured in some $\delta$ Scuti stars 
($\theta$ Tuc: \citet{ps96}; XX Pyx: \citet{hand00}).  The small separations 
are also a possible tool, though they are very dependent upon the mode 
identification, and it is unlikely that modes of $\ell \geq 4$ will be 
detected photometrically, limiting the 
usefulness of this quantity (since the only the small separations for 
$\ell = 0$ or $1$ may be calculated in this case).

Another diagnostic of the interior is the frequency difference, specifically
the second and fourth differences, defined by

\begin{equation}
\delta^{2} = \nu_{n+1} - 2 \nu_{n} + \nu_{n-1}
\end{equation}

\begin{equation}
\delta^{4} = \nu_{n+2} - 4 \nu_{n+1} + 6 \nu_{n} - 4 \nu_{n-1} + \nu_{n-2}
\end{equation}

\noindent
These are useful for detecting discontinuities in the radial sound speed gradient.
These discontinuities cause an ``oscillation'' in the frequency difference,
with a frequency corresponding to the sound crossing time between the surface 
and the discontinuity.  There are several problems with using this quantity in
$\delta$ Scuti stars.  First, it requires that the radial mode order $n$ be 
known.  If the observed stellar pulsation spectrum is free of gaps (as
in the Sun), this is straightforward, but again $\delta$ Scuti stars have
never exhibited all of the theoretically predicted modes at once.  Second,
rotational splitting of modes introduces an additional complication to the
differences, since the equations above assume $m = 0$, and that there is no
second-order rotational perturbation of the spectrum.  Again, this is unlikely
to be the case in $\delta$ Scuti stars.  Third, $\delta$ Scuti stars away from
the red edge of the instability strip have small shallow surface convection
zones, rather than the thick convection zone we see in the Sun.  $\delta$ Scuti
stars may have a core convection zone, but the effect of this on the frequency
differences in unclear, particularly since the deep layers are better probed
by the $g$- and mixed-modes which do not follow the above equation as do the
$p$-modes.  We tested a the second- and fourth-differences for several hundred
pulsation models along a single track, and found no easily detectable
seismological signature, other than the large spike at the acoustic cutoff
frequency.

\section{FG Vir -- a case study}
Now, we turn from the general behavior of $\delta$ Scuti stars to a more
specific case, namely FG Virginis.  FG Vir ($m_{V} = 6.56$) is a well-studied 
$\delta$ Scuti star, in which at least two dozen pulsation frequencies have 
been detected \citep{breger98,B2k}.
Several of these frequencies have mode identifications, obtained by measuring
the temporal phase shifts in different Str\"{o}mgren filters 
\citep{bppg99,ps00}. The detection of multiple modes along
with mode identifications raises the real possibility of performing rigorous
asteroseimology on stars other than the Sun.  Breger et al.~found that two of
the modes observed in FG Vir are likely radial ones, corresponding to the
fundamental and third radial overtone.  (They claim the higher frequency may
be $\ell = 1$, but for the purpose of this work, we assume the $\ell = 0$
identification is the correct one.)  We can use these two modes to constrain
the parameter space of models of FG Vir.

The major difficulty in performing asteroseismology by ``forward'' modeling
is in constraining the parameter space of models.  There are several 
variables which must be closely estimated when building a model: mass,
temperature, luminosity, chemical composition (both metals and helium),
core overshooting length, and age are all important first-order parameters.
Other parameters such as rotation, diffusion, and mixing length are also
important and further complicate the modeling.  \citet{bppg99} overcame
this by generating a large grid of pulsation models consistent with the 
observed properties of FG Vir ($\log{g} = 4.00 \pm 0.1$, 
$\log{T} = 3.875 \pm 0.006$).  They first constructed a grid of more than a 
dozen evolution models and generated pulsation models spaced evenly along
each evolution track.  This yielded hundreds of candidate models from which
a test model was obtained.  \citet{pam98} attempted a similar fitting
process, generated 40,000 models, and minimized the $\chi^{2}$ of the 
(observed $-$ calculated) frequency residuals.  In their case, they were able
to obtain close matches to XX Pyx, but no exact fits.  The difficulty with 
both of these methods is that they require the computation of many thousands
of evolution and pulsation models, with a wide range of stellar parameters.
As \citet{pam98} note, even when many thousands of models are produced, there
is no guarantee that any will be an exact match to a given star, and that
even the best-fitting model from a given set can likely be improved by
modifying the chemical composition, overshooting parameter, and rotation
profile.  Therefore, it is important to study not just individual stars, but
the behavior of {\it models} as functions of input parameters.

In this study, we adjust the important parameters
{\it individually}, to see how we might be able to more easily constrain the 
parameter space of models to obtain close matches to the observed
star.  In particular, we wish to show how changing individual parameters
affects a few observable quantities rather than the entire pulsation spectrum.
We note that we will not attempt to ``fit'' a model to FG Vir.  Such a process
would require the generation of thousands of pulsation models, and the 
caluclation of $\chi^{2}$ based upon the difference between observed and
calculated frequencies \citep{pam98}.  What we are doing
here is determining whether we can use the model behavior in the period-period 
ratio and temperature-period ratio planes to {\it constrain} the parameter
space of models, in hopes that the number of models required for brute-force
fitting might be significantly reduced.

For our tests, we study how changes to the mass, helium abundance, metal
abundance, and convective core mixing length individually affect the evolution
of the star.  However, we look at these changes not in the traditional
temperature-luminosity plane, but in two other observable planes: the
fundamental period-period ratio plane and the temperature-period ratio plane.
We use this information to search for models with the desired fundamental 
period, period ratio, and effective temperature.  First, we generated a
grid of evolution models, shown in Table \ref{tab:two}, designed to cover 
the likely parameters of FG Vir.  For each set of models, we change only
one parameter at a time, so we can test the effects of changing each parameter 
individually.

\begin{table}[htbp]
\caption{Evolution model initial conditions for the FG Virginis grid. 
$Y_{0}$ and $Z_{0}$
are the initial helium and metal mass fractions, and $\alpha_{C}$ is the
convective core overshooting parameter in pressure scale heights.  Models
without overshoot are labeled ``none''.}
\label{tab:two}
\begin{center}
\begin{tabular}{cccc|cccc}\hline
$Z_{0}$ & $Y_{0}$ & Mass & $\alpha_{C}$ & $Z_{0}$ & $Y_{0}$ & Mass & 
$\alpha_{C}$ \\
\hline \hline
0.02 & 0.28 & 1.9 & none  & 0.01 & 0.28 & 1.9 & none\\
0.02 & 0.28 & 2.0 & none  & 0.02 & 0.28 & 1.9 & none\\
0.02 & 0.28 & 2.1 & none  & 0.03 & 0.28 & 1.9 & none\\
\\
0.02 & 0.26 & 1.9 & none  & 0.02 & 0.28 & 1.8 & none\\
0.02 & 0.28 & 1.9 & none  & 0.02 & 0.28 & 1.8 & 0.1 \\
0.02 & 0.30 & 1.9 & none  & 0.02 & 0.28 & 1.8 & 0.2 \\
     &      &     &       & 0.02 & 0.28 & 1.8 & 0.3 \\
\hline
\end{tabular}
\end{center}
\end{table}

\subsection{Results}
\begin{figure}
\plotone{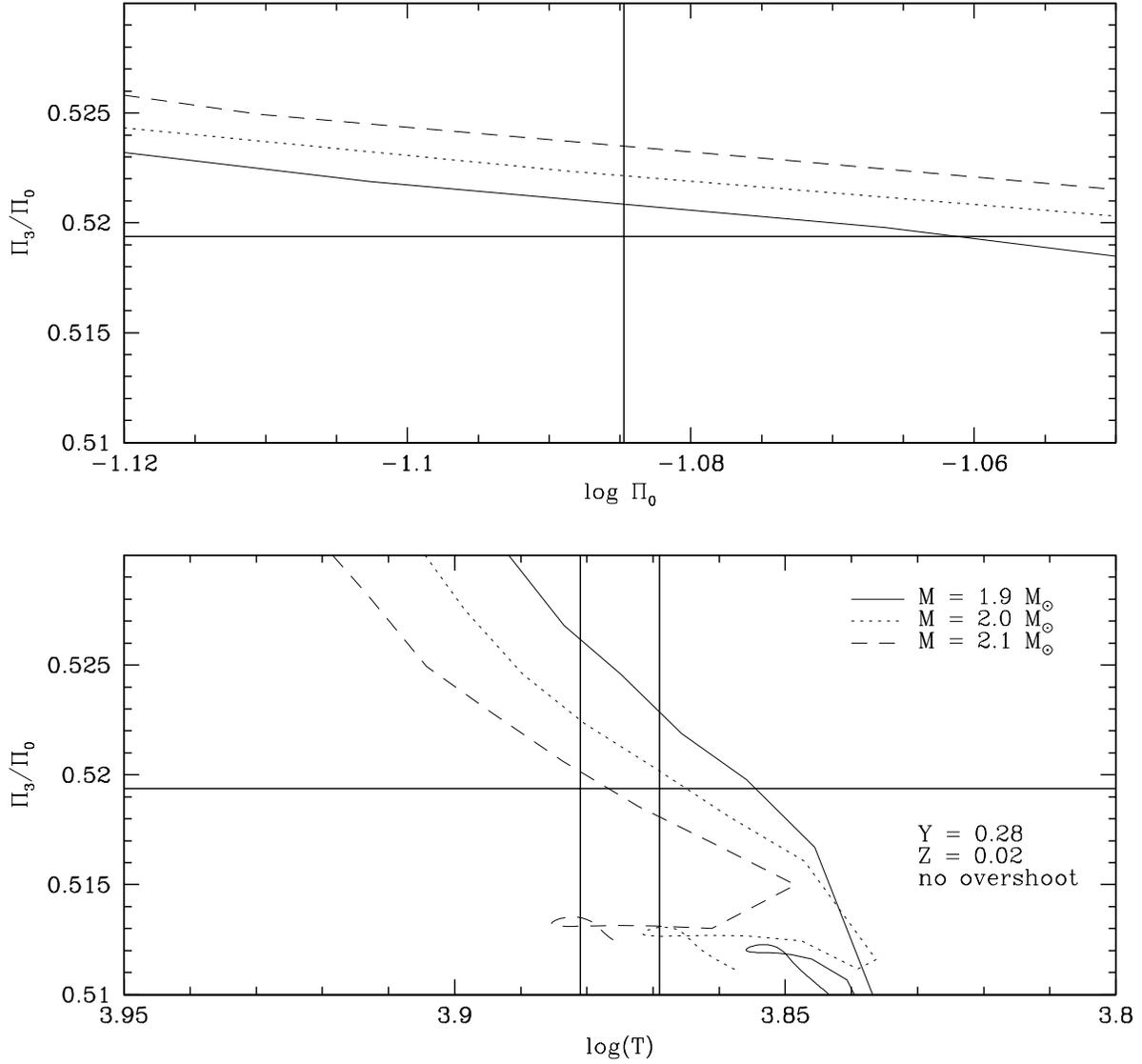}
\caption{Evolution of FG Virginis models with different masses in the
period-period ratio (top) and temperature-period ratio (bottom) planes.
The cross hairs in each box denote the observational data or constraints
for FG Vir.  All models have Z=0.02, Y=0.28, and no core overshooting.
Solid lines - M=1.9M$_{\odot}$; dotted lines - M=2.0M$_{\odot}$; dashed
lines - M=2.1M$_{\odot}$.}
\label{fig:eight}
\end{figure}

We find that changes to three of the parameters -- mass, and the helium
and metal abundances -- affect the models in a similar way; adjusting these
parameters moves the model to the desired location in the period-period ratio
plane, but away from the desired location in the temperature-period ratio
plane and vice versa.  This can be seen in Figure \ref{fig:eight}, where we show how the
models shift in the two planes as we change the mass; none of 
the test models passed through the desired points in either plane, and 
adjustment of one single parameter worsened the fit in at least one of the 
planes.  The effects of changing the chemical abundances are similar to those
of changing the mass.  Changes in the metal abundance produced the largest 
effect, but the steps in the metal abundance were very large ($\sim 50 \%$).  
Changes in the mass also had a significant effect.  Models
appeared to be less sensitive to changes in the helium abundance.

\begin{figure}
\plotone{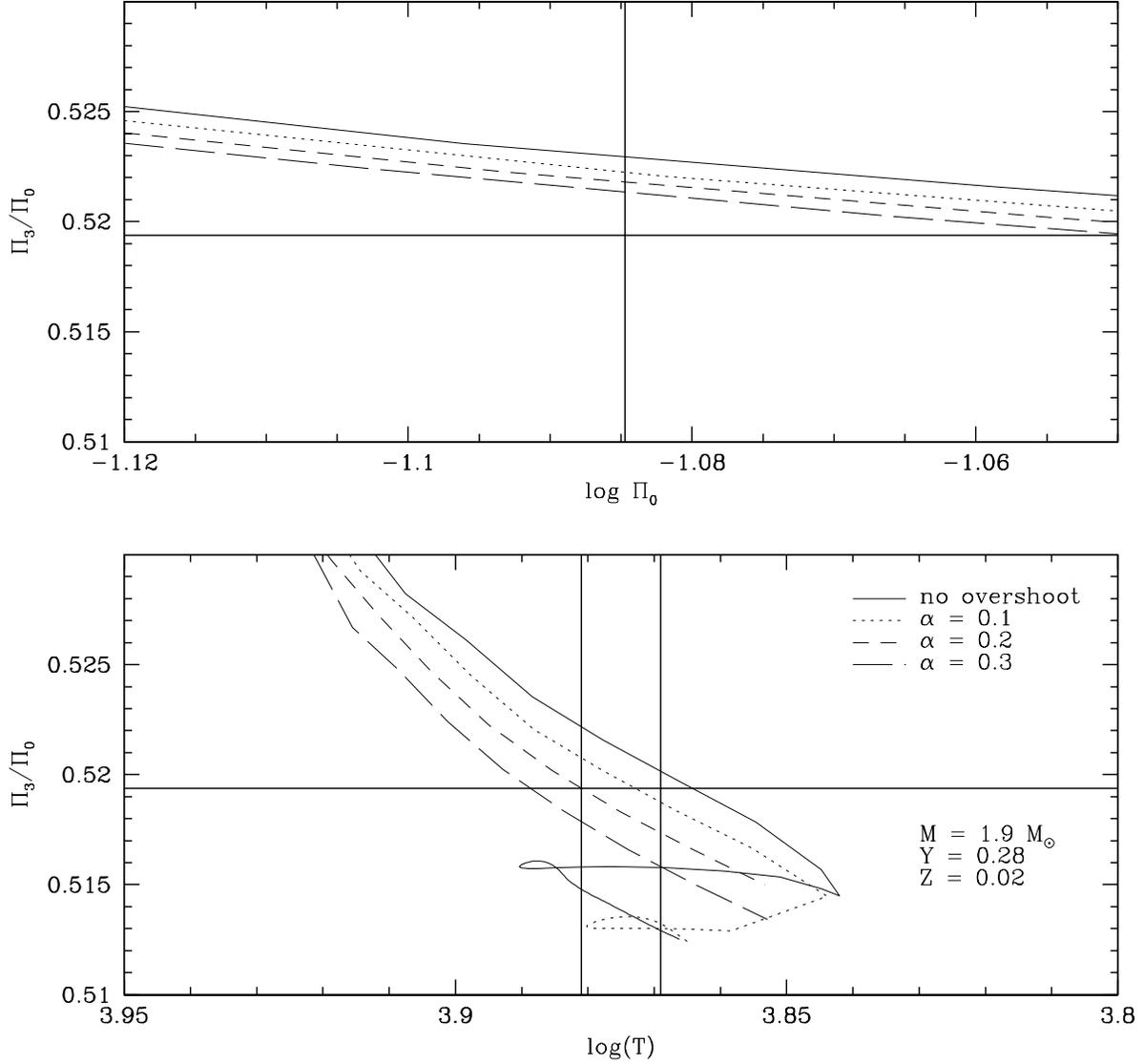}
\caption{Evolution of FG Virginis models with different core overshooting
parameters in the period-period ratio (top) and temperature-period ratio
(bottom) planes.  The cross hairs in each box denote the observational data or
constraints for FG Vir.  All models have Z=0.02, Y=0.28, and
M=1.9M$_{\odot}$.  Solid lines - $\alpha_{C}=0.1$; dotted lines -
$\alpha_{C}=0.2$; dashed lines - $\alpha_{C}=0.3$.  Note that unlike changes
to the mass, and to helium and metal abundances, increasing the convective 
core overshooting parameter moves the evolution tracks toward the desired 
locations in both planes.}
\label{fig:nine}
\end{figure}

The situation is different for variation of the convective core overshooting
parameter, shown in Figure \ref{fig:nine}.  In this case, as the core overshooting value
is increased from zero (no overshoot) to 0.3 (substantial overshoot), models
move closer to the desired loci in both the period-period ratio and 
temperature-period ratio planes simultaneously.  This is significant because
it suggests that one can tune the mass and chemical abundances to produce a
best ``near-fit'', and then vary the core overshooting parameter to obtain
a true best-fit.  The reason for this is that the effect of changing the
overshooting parameter on the evolution model is fundamentally different than
a change in the mass or chemical composition.  A change in the mass or
chemical composition translates the evolution model to a different 
location in the temperature-luminosity diagram.  However, a change in the
overshooting parameter changes the stellar radius and mean
density over time, without a change in temperature.

\subsection{A test model}
\begin{figure}
\plotone{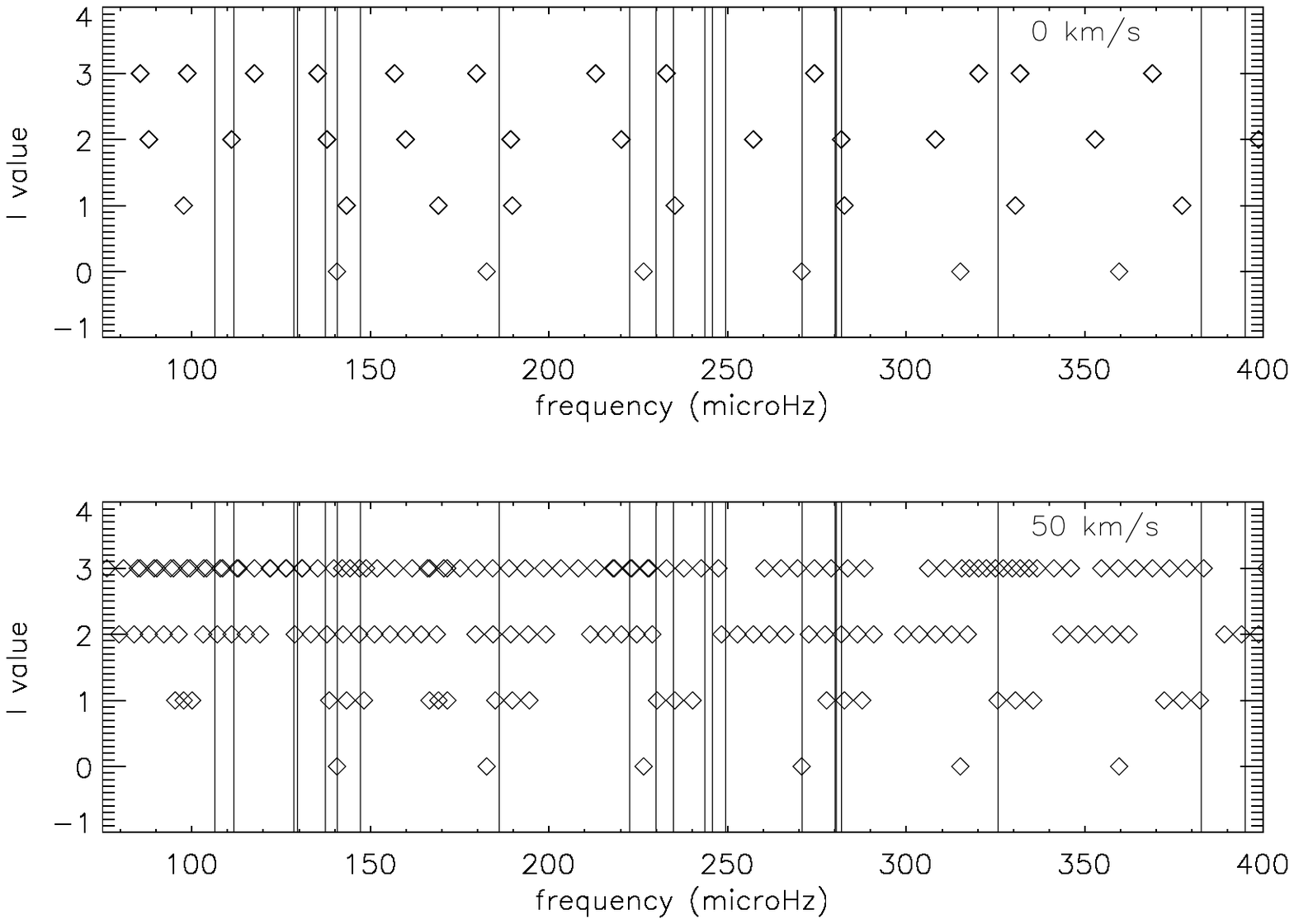}
\caption{Best-fit model for FG Virginis.  The top panel shows the observed
pulsation modes of FG Vir (vertical lines) with the theoretically predicted
$\ell = 0,1,2$ and $3$ modes of this particular model.  Several observed
modes do not have close matches at any $\ell$-value.  The bottom panel shows
the same spectrum but with first-order rotational splittings calculated for
a rotation velocity of 50 km/s.  It is clear that $\ell = 3$ modes can fit
nearly any mode at that rotation velocity, though matches were obtained 
from among the $\ell = 1,2$ modes as well.  The radial modes are unaffected 
in first-order rotational splitting.}
\label{fig:ten}
\end{figure}

We present a model closely matching the frequencies of FG Vir obtained using 
the models of the previous
section as a guide.  We selected a mass and chemical composition so as to
minimize the distance from the model evolution tracks to the desired location
in the observational planes, and then adjusted the core overshooting length
to obtain a reasonable fit.  The model shown has 
$(X,Y,Z) = (0.69,0.28,0.03)$, $M = 1.9M_{\odot}$, $\log{T}= 3.870$, 
$\log{L}= 1.151$,
and a convective core overshooting parameter $\alpha_{C} = 0.3$, at an age
of 0.93 Gyr.  We present the $\ell = 0, 1, 2,$ and $3$ model frequencies for
both a non-rotating case and a model with first-order rotational splittings
of 50 km/s in Figure \ref{fig:ten}.  The modes corresponding to the identified 
modes in \citet{bppg99} are shown in Table \ref{tab:three}.

\begin{table}[htbp]
\caption{Theoretical fits to the observed spectrum of FG Virginis.  Models
without rotation ($v=0$) and with first-order rotational splitting ($v=50$)
are shown.  The model with rotational splitting can fit most of the observed
modes having identifiable $\ell$-values.}
\label{tab:three}
\begin{center}
\begin{tabular}{cccccccc}\hline
$\nu_{obs}$ & $\ell_{obs}$ & $\nu_{v=0}$ & $\ell_{v=0}$ & $\nu_{v=50}$ & 
$\ell_{v=50}$ & $\delta\nu_{v=50}$ & $m_{v=50}$ \\
\hline \hline
140.67 & 0      & 140.64 & 0 & 140.64 & 0 &       &    \\
270.87 & 0 or 1 & 270.70 & 0 & 270.70 & 0 &       &    \\
106.47 & 2      &  none  &   & 111.12 & 2 &  3.98 & -1 \\
111.76 & 1 or 2 & 111.12 & 2 & 111.12 & 2 &       &    \\
147.18 & 1      & 143.32 & 1 & 143.32 & 1 &       &    \\
229.95 & 2      &  none  &   & 235.21 & 1 &  4.98 & -1 \\
243.66 & 2      &  none  &   & 257.20 & 2 &  4.46 & -2 \\
280.42 & 1 or 2 & 281.80 & 2 & 281.80 & 2 &       &    \\
\hline
\end{tabular}
\end{center}
\end{table}

As Table \ref{tab:three} and Figure \ref{fig:ten} show, the non-rotating 
model is not a
good match of the observed frequencies, either in general or using the
mode identifications of Breger et al.  In particular, there are several
closely-spaced modes which do not have corresponding pairs and triplets in
the theoretical spectrum.  Three of the observationally identified modes
do not have any corresponding theoretical mode, suggesting that the 
nonrotating model is a poor match.  The model with first-order rotational
splitting is a somewhat better
match, since splitting of order 4-5 microhertz can match some modes
not detected in the non-rotating case.  The worst match is the mode at
243.66 $\mu$Hz which requires the $m = -2$ split, and which is still too high
by more than four microhertz.  A higher rotational velocity ($v > 60$ km/s) 
would
improve the matches slightly.  If we consider the $\ell = 3$ modes, nearly 
every mode could be fit by a theoretical counterpart.  However, no mode 
identifications with $\ell > 2$ have been made, and it is debatable whether
$\ell \geq 3$ modes are detectable in the integrated light of a star.  
Furthermore, if the observational mode identifications are not followed, modes
of $\ell \leq 2$ can fit nearly every observed mode.

It is important to note that the evolution models were computed without 
including the effects of rotation, and that the first order rotational
splittings are not sufficient for the case of rapid rotation.  Second-order
rotational distortion of the star changes the radial mode frequencies in
addition to the nonradial, $|m| > 0$ modes \citep{dg92,tbg00}.
This significantly complicates this
analysis method if the rotation rate is not known, but in principle the
effect of the rotation rate on evolution tracks in the period-period ratio 
and temperature-period ratio planes could also be calculated.  Our current
pulsation code does not include the second-order effects of rotation, so we
leave this analysis for a future paper.  However, the model chosen using the 
method outlined above with only first-order rotational splitting provides 
a reasonable theoretical match to the observed modes.
The model shown above is not a closer match to FG Vir than any other models
computed thus far \citep{bppg99,gbt00}, however, the fact that we were able
to find a reasonable match to observations using only the identification of
two modes and the constraints on effective temperature suggests that this
method may be a reasonable way to effectively constrain the parameter space
of models when attempting a true asteroseismological fit.

\section{Conclusions}
We have shown that modifying the fundamental stellar parameters of mass, 
chemical composition, and core overshoot parameter can affect the evolution
and pulsation in a significant way.  We have varied the mass, metal and helium
abundances, and the convective core overshooting parameter individually to
test the consequences on the evolution of the stellar model and its pulsation
spectrum.  We have found that varying the mass has the largest effect on the
pulsation frequencies, with significant shifts to both $p$- and $g$-modes.  
Varying the metal abundance has a significant effect on the $p$-modes, and less
on the $g$-modes.  Varying the helium abundance has very little effect on the
pulsation spectrum, despite having a drastic effect on effective temperature
and luminosity.  Varying the convective core overshooting parameter also has
an effect on the frequencies, though unlike the effects of the other three 
parameters, the
effects strongly increases with time, with no difference observed at the
ZAMS (as expected).  We find that the post-main sequence evolution of 
overshooting models is relatively more rapid than in non-overshooting models.
We suggest that this may provide a way to calibrate the overshooting parameter
using period changes.  We also discuss a few other possible diagnostics.
The large separations are probably the best
diagnostic at the current time given that not all predicted modes are excited
in these stars at once. Better photometry from space-based missions may make
the other quantities useful as well.  The second and fourth
differences do not appear to be useful for studying stellar convection in
$\delta$ Scuti stars as they are in the Sun, though a sensitive analysis has
not yet been completed.  The seismological signatures in the small separations
and the second and fourth differences will also be contaminated heavily by
rotational splitting, and by the presence of $g$- and mixed-modes in the
spectrum.

Finally, we have shown how adjustment of individual stellar parameters affects 
the pulsation behavior for the specific case of FG Virginis models. 
To do this, we traced the evolution of the star in the period-period ratio and 
temperature-period ratio planes, using only two of the twenty-two observed
pulsation modes and the effective temperature as observational constraints.
The evolution of models in these planes is
fully consistent with those shown in \citet{bppg99}, though in our case, we
were able to match the observed period and period ratio with a model
computed using standard physical data.
We found that varying the convective core overshooting parameter changes the 
pulsation behavior in a fundamentally different manner than do changes to the 
mass and chemical composition.  We use this information to attempt to model FG
Virginis.  A close match to the observed pulsation spectrum was not obtained,
though a model with $M = 1.9 M_{\odot}$, $Z = 0.03$, $Y = 0.28$, and 
$\alpha_{C} = 0.3$ can match most modes if first-order rotational splitting 
of order 50 km/s is applied.  Second-order rotational effects must be 
included in the pulsation analysis to obtain a truly realistic model, since 
the rotation rate is high enough to significantly distort the mode 
frequencies, and will induce coupling between the $\ell = 0,2$ and 
$\ell = 1,3$ modes.  Further refinement of our modeling 
method is required, particularly to include the effects of rotation as a 
fifth adjustable parameter.

\clearpage

\begin{thebibliography}{}
\bibitem[Alexander(1995)]{a95} Alexander, D.R., 1995, in {\it Astrophysical
Applications of Powerful New Databases}, edited by S.J. Adelman and
W.L. Weise, ASP Conf. Ser. 78, 63
\bibitem[Basu(1997)]{basu97} Basu, S., 1997, \mnras~288, 572
\bibitem[Breger(2000)]{B2k} Breger, M., 2000, in {\it Delta Scuti and Related
Stars}, ASP Conf. Ser. 210, 3
\bibitem[Breger et al.(1998)]{breger98} Breger, M. et al., 1998, \aap~331, 271
\bibitem[Breger et al.(1999)]{bppg99} Breger, M. et al., 1999, \aap~341, 151
\bibitem[Breger et al.(1999b)]{breg4cvn} Breger, M. et al., 1999, \aap~349, 225
\bibitem[Chaboyer, Demarque, \& Pinsonneault(1995)]{yrec} Chaboyer, B.,
Demarque, P., \& Pinsonneault, M.H., 1995, \apj~441, 865
\bibitem[Christensen-Dalsgaard(2000)]{cd2000} Christensen-Dalsgaard, J., 2000,
in {\it Delta Scuti and Related Stars}, ASP Conf. Ser. 210, 187
\bibitem[Cox, King, \& Hodson(1979)]{ckh79} Cox, A.N., King, D.S., \& Hodson,
S.W., 1979, \apj~228, 870
\bibitem[Dziembowski \& Goode(1992)]{dg92} Dziembowski, W.A. \& Goode, P.R., 
1992, \apj~394, 670
\bibitem[Dziembowski \& Krolikowska(1990)]{dk90} Dziembowski, W. \& 
Krolikowska, M., 1990, {\it Acta Astronomica} 40, 19
\bibitem[Gough(1990)]{gough90} Gough, D.O., 1990, in {\it Progress of Seismology
of the Sun and Stars}, Lecture Notes in Physics 367, 283
\bibitem[Gough \& Novotny(1990)]{gn90} Gough, D.O. \& Novotny, E., 1990, 
\solphys~128, 143
\bibitem[Grevesse \& Noels(1993)]{gn93} Grevesse, N. \& Noels, A., 1993, in
{\it Origin and Evolution of the Elements}, edited by M. Prantzos, 
E. Vangioni-Flam and M. Casse, Cambridge University Press
\bibitem[Guzik et al.(2000)]{gbt00} Guzik, J.A., Bradley, P.A., \& Templeton,
M.R., 2000, in {\it Delta Scuti and Related Stars}, edited by M. Breger and
M. Montgomery, ASP Conf. Ser. 210, 247
\bibitem[Handler et al.(2000)]{hand00} Handler, G., et al., 2000,
\mnras~318, 511
\bibitem[Harrison et al.(2000)]{har00} Harrison, T.E., et al., 2000, 
\aj~120, 2649
\bibitem[Iglesias \& Rogers(1996)]{opalo} Iglesias, C.A. \& Rogers, F.J., 1996,
\apj~464, 943
\bibitem[Michel et al.(1998)]{mich98} Michel, E., et al., 1998, in {\it A
Half Century of Stellar Pulsation Interpretation: A Tribute to Arthur N. Cox},
edited by P.A. Bradley and J.A. Guzik, ASP Conf. Ser. 135, 475
\bibitem[Napiwotzki, Sch\"{o}nberner, \& Wenske(1993)]{nsw93} Napiwotzki, R.,
Sch\"{o}enberner, D., \& Wenske, V., 1993, \aap~268, 653
\bibitem[Osaki(1975)]{osaki75} Osaki, Y., 1975, \pasj~27, 237
\bibitem[Pamyatnkyh et al.(1998)]{pam98} Pamyatnykh, A.A. et al., 1998,
\aap~333, 141
\bibitem[Paparo et al.(1996)]{ps96} Paparo, M. et al., 1996, \aap~315, 400
\bibitem[Paparo \& Sterken(2000)]{ps00} Paparo, M. \& Sterken, C., 2000, 
\apj~362, 245
\bibitem[Perryman et al.(1997)]{hip97} Perryman, M.A.C., et al., 1997, \aap~323,
L49
\bibitem[Petersen(1973)]{pet73} Petersen, J.O., 1973, \aap~27, 89
\bibitem[Petersen(1993)]{pet93} Petersen, J.O., 1993, \apss~210, 153
\bibitem[Rogers, Swenson, \& Iglesias(1996)]{opale} Rogers, F.J., Swenson, F.J.,
\& Iglesias, C.A., 1996, \apj~456, 902
\bibitem[Solano \& Fernley(1997)]{sf97} Solano, E. \& Fernley, J., 1997, 
\aaps~122, 131
\bibitem[Soufi, Goupil, \& Dziembowski(1998)]{sgd98} Soufi, F., Goupil, M.J.,
\& Dziembowski, W.A., 1998, \aap~334, 911
\bibitem[Templeton(2001)]{tem01} Templeton, M.R., 2001, \apj~submitted
\bibitem[Templeton, Bradley, \& Guzik(2000)]{tbg00} Templeton, M.R., Bradley,
P.A., \& Guzik, J.A., 2000, \apj~528, 979
\bibitem[Viskum(1997)]{viskum97} Viskum, M., 1997, PhD Thesis, Aarhus 
Universiteit
\end{thebibliography}
\end{document}